\documentclass[twocolumn,11pt]{article}

\usepackage{graphicx}

\newcommand{\be}{\begin{equation}}
\newcommand{\ee}{\end{equation}}
\renewcommand{\k}[2]{\frac{#1}{#2}}

\def\p{\partial}
\def\s{\,\,\,\,}
\def\bb{\beta}
\def\b{\bar \beta}
\def\0{^{(0)}}
\def\1{^{(1)}}

\def\={\approx}

\def\v{\lambda}
\def\ra{\rightarrow}
\newcommand{\pab}[2]{\frac{\p #1}{\p #2}}

\title{Pressure Solution in Sedimentary Basins: \\ Effect of
Temperature Gradient}

\author{Xin-She Yang     \\
  Department of Fuel and Energy and
  Department of Applied Mathematics\\
  University of Leeds, LEEDS LS2 9JT, UK }

\date{}

\begin{document}

\maketitle

\abstract{\small 

Pressure solution is an important
process in sedimentary basins, and its behaviour depends mainly
on the sediment rheology and temperature distribution.
The compaction relation of pressure solution is typically
assumed to be a viscous one and is often written as a relationship
between effective stress and strain rate.
A new derivation of viscous compaction relation is formulated
based on more realistic boundary conditions at grain contacts.
A nonlinear diffusion problem with a moving boundary is solved
numerically and a simple asymptotic solution is given to
compare with numerical simulations. Pressure solution is
significantly influenced by the temperature gradient.
Porosity reduction due to pressure solution is enhanced
in an environment with a higher thermal gradient, while
porosity decreases much slowly in the region where the thermal
gradient is small. Pressure solution tends to complete
more quickly at shallower depths and earlier time in higher
temperature environment than that in a low one. These features
of pressure solution in porous sediments are analysed
using a perturbation method to get a solution for the steady
state.  Comparison with real data shows a reasonably very good
agreement.     \\


}

\noindent {\bf Citation detail}: X. S. Yang, Pressure solution in sedimentary basins: effect of temperature
gradient, {\it Earth and Planetary Science Letters}, {\bf 176}, 233-243 (2000).

\section{INTRODUCTION}

Pressure solution is a very common and important deformation process
in porous media and  granular materials such as sediments and soils.
Pressure solution also occurs in
sedimentary basins where hydrocarbons and oil are primarily formed.
The modelling of such compactional flow is thus important in the
oil industry as well as in civil engineering.
One particular problem which affects
drilling process is the occasional occurrence of abnormally
high pore fluid pressures, which, if encountered suddenly,
can cause drill hole collapse and consequent failure of the
drilling operation.   Therefore, an industrially important
objective is to predict overpressuring before drilling and
to identify its precursors during drilling.
An essential step to achieve such objectives
is the scientific understanding of their mechanisms and the
evolutionary history of post-depositional sediments such as shales.

Compaction is the process of volume reduction via
pore-water expulsion within sediments due to the increasing weight of
overburden load. The requirement of its occurrence is not only the
application of an overburden load but also the expulsion of pore water.
The extent of compaction is strongly influenced by sedimentation history and
the lithology of sediments. The freshly deposited loosely packed
sediments tend to evolve, like an open system, towards a closely packed
grain framework during the initial stages of burial compaction and this is
accomplished by the processes of grain slippage, rotation, bending and
brittle fracturing. Such reorientation processes are collectively
referred to as  mechanical compaction,
which generally takes place in  the first 1 - 2 km of burial.
After this initial porosity loss,
further porosity reduction is accomplished by the process of
 chemical compaction such as pressure solution at grain contacts [1,2,3].

Pressure solution has been considered as an important
process in deformation and porosity change during compaction
in sedimentary rocks [4,5]. Pressure solution refers
to a process by which grains dissolve at
intergranular contacts under non-hydrostatic stress
and reprecipitate in pore spaces, thus resulting in compaction.
The solubility of minerals increases with increasing effective stress at
grain contacts. Pressure dissolution at grain contacts is therefore a
compactional response of the sediment during burial in an attempt to
increase the grain contact area so as to distribute the effective
stress over a larger surface.  Such a compaction process
is typically assumed to viscous [5,6,7] and it is usually
referred to as viscous compaction, viscous creep or pressure
solution creep. Its rheological constitutive relation
(or compaction relation) is often written as a relationship
between effective stress and strain rate.

A typical form of pressure solution is
intergranular pressure solution (IPS) which occurs
at individual grain contacts  and free face
pressure solution (FFPS) which occurs at the
face in contact with the pore fluid, but most studies
have concentrated on the former one (IPS).
Extensive studies [1,5,6,7,8] on pressure solutin
have been carried out in the last two decades, and a
comprehensive literature review on these models  was given by
Tada and Siever [8]. A more recent and brief review can be found in [5,6].
Despite of its geological importance,
the mechanism of pressure solution is still poorly understood.
Recently, Fowler and Yang [5] present a new
mathematical approach to model pressure solution and viscous compaction
in sedimentary basins and show that the main parameter controlling the
compaction and porosity reduction is the compaction parameter $\v$, the
ratio of hydraulic conductivity to the sedimentation rate.   Compaction
relation of poro-elastic and viscous type is also an important factor
controlling the behaviour of compaction profile. However, the temperature
effect has not been included in their approach. Thus, we mainly investigate
the effect of different temperature gradients on the viscous compaction
due to pressure solution in sedimentary basins.

\section{MATHEMATICAL MODEL}

For the convenience of investigating the effect of  compaction in porous
media  due to pure density differences, we will assume the basic model
of compaction is rather analogous to the process of soil consolidation.
The porous  media act as a compressible porous matrix, so that mass
conservation of pore fluid together with  Darcy's law leads to the
1-D model equations of the general type [1,5].

\begin{equation}
\pab{\rho_s (1-\phi)}{t} +\pab{}{z}[\rho_s (1-\phi) u^{s} ]=0,
\s ({\rm solid \,\, phase})
\label{VC:MASS}
\end{equation}
\begin{equation}
\pab{\rho_l \phi}{t}+ \pab{(\rho_l \phi  u^{l})}{z}=0,
\s ({\rm liquid \,\, phase})
\end{equation}
\begin{equation}
\phi (u^{l}-u^{s})
=-\frac{k(\phi)}{\mu}[ \pab{p}{z}+\rho_l g ],
\s ({\rm Darcy's \,\, law})
\end{equation}
\[
-\pab{}{z}[(1+\k{4 \eta}{3 \xi}) p_e]
-\pab{p}{z}-[\rho_s (1-\phi)+\rho_l \phi] g=0, \]
\be ({\rm force \,\, balance}),
\ee
where  $u^{l}$ and $u^{s}$ are the velocities of fluid and
solid matrix, $k$ and $\mu$ are the matrix permeability and
the liquid viscosity,   $\rho_{l} $ and $\rho_{s}$ are the
densities of fluid and solid matrix, $p$ is pore pressure,
$p_{e}$ is the effective pressure, $\eta$ is medium viscosity and
$\xi$ is compaction viscosity, and $g$ is the gravitational acceleration.
Combining the force balance and Darcy's law to eliminate $p$, we have
\begin{equation}
\phi (u^{l}-u^{s})
=\frac{k(\phi)}{\mu} \{ \pab{}{z}[(1+\k{4 \eta}{3 \xi}) p_e]-(\rho_s-\rho_l)(1-\phi) g \},
\end{equation}
which is a derived form of Darcy's law. By assuming the densities
$\rho_s$ and $\rho_l$ are constants, we can see that only the density
difference $\rho_s-\rho_l$ is important to the flow evolution. Thus,
the compactional flow is essentially density-driven flow in porous
media.

Compaction relation is a relationship between effective pressure $p_e$ and
strain rate $\dot e=\pab{u^s}{z}$ or porosity $\phi$ [5,6,7].
The common approach
in soil mechanics and sediment compaction is to model this generally nonlinear
behaviour as  poroelastic, that is to say, a relationship of Athy's law type
$p_e = p_e (\phi)$, which is derived from fitting the real data of sediments.
However, this
poroelastic compaction law is  only valid for the  compaction in porous media
in the upper and shallow region, where compaction  occurs due to the pure
mechanical movements such as  grain sliding and packing rearrangement. In the
more deeper region, mechanical compaction is gradually replaced by the
chemical compaction due to stress-enhanced flow along the grain boundary
from the grain contact areas to the free pore, where pressure is essentially
pore pressure. A typical process of such chemical compaction in sediment
is pressure solution whose rheological behavior is usually viscous, so that
it sometimes called viscous pressure solution.

The mathematical formulation for viscous compaction is to derive a relation
between creep rate $\dot e$ and effective stress $\sigma_e$. Rutter's
creep relation is widely used [7,8,9]
\begin{equation}
\dot e=\frac{A_{k} c_{0} \, w D_{gb} }{\rho_{s} \bar d^{3}}
\sigma_e, \label{CREEP-1}
\end{equation}
where $\sigma_e$ is the effective normal stress across the grain contacts,
$A_{k}$ is a constant, $c_{0}$ is the equilibrium concentration
(of quartz) in pore fluid,  $\rho, \, \bar d$ are the density
and (averaged) grain diameter (of quartz). $D_{gb}$ is the diffusivity
of the solute in water along grain boundaries with a thickness $w$.
$D_{gb}$ also varies with temperature $T$
\be
D_{gb}(T) =D_{gb} e^{-\frac{E_a}{R T}},  \label{CREEP-11}
\ee
where $E_a$ is the effective activation energy with a value of
$3 \sim 6$ kJ/mole or even much lower [1, 6]. From the values of the
diffusion coefficient in quartz-water and
rocksalt-water systems at $300, \, 600, 1200$ K, we get an estimate value
of $E_a \approx 0.65$ kcal/mole [1,7].

Note that $\sigma_e=-(1+\k{4 \eta}{3 \xi}) p_e$ and $\dot e=\pab{u^{s}}{z}$.
With this, (\ref{CREEP-1}) becomes the following  compaction law
\begin{equation}
p_{e}=-\xi \nabla . {\bf u}^{s}.   \label{VC:CREEPNEW}
\end{equation}
More generally speaking, $\xi$ is also a function of porosity $\phi$.
The compaction law is analogous to
Fowler's viscous compaction laws used in studies of
magma transport in the Earth's mantle.

\subsection{Derivation of Viscous Law}

The approach of deriving the law of viscous compaction depends on the
underlying mechanism. The classical theoretical consideration  assumed
a grain-boundary diffusion film
of constant thickness and diffusivity, while others used the concept
of a roughened, fluid-invaded non-equilibrium contact structure
Shimizu [6] presented a kinetic approach extending
Coble's classical treatment of grain boundary diffusion creep
by including the kinetics of quartz dissolution/precipitation reaction.
Shimuzu's  derivation is instructive although the boundary
conditions used in his formulation are questionable
and unrealistic. In addition, Shimuzu's 1-D
approximation is only valid for a {\em closed
system} due to $\pab{c}{x}=0$ used in his work
when the thickness $w$ of the water film is
small with respect to the grain diameter ($\bar d$) [5,10].
In order to correctly formulate the derivation,
we now provide a new derivation by using more
realistic boundary conditions in an {\it open
system}.

Now let us consider the intergranular contact region as a disk with a
radius $r=L$. Let $J(r)$ be the radial component of solute mass flux,
$\dot e$ be the average strain rate, and $v$ is the
uniform shortening velocity of the upper grain relative to the lower
grain due to the pressure solution creep [6,10]. The kinetic relation
between $v$ and $\dot e$ becomes
\begin{equation}
v=\dot e \bar d.  \label{V:DOTE}
\end{equation}
For simplicity, we assume that the film thickness $w$ is constant
and the diffusion is near steady-state. Mass conservation gives
\begin{equation}
2 \pi r J(r) + \rho_{s} \pi r^{2} v=0,
\end{equation}
where the flux $J(r)$ obeys {\em Fick's Law}
\begin{equation}
J(r)=-D_{gb} w \frac{dc}{dr}.
\end{equation}
The steady-state solution of concentration $c(r)$ for the
boundary conditions $c_{r}=0$ at $r=0$, $c=c_{0}$ at $r=L$ is
\begin{equation}
c(r)=c_{0}-\frac{\rho_{s} v}{4 D_{gb} w} (L^{2}-r^{2}).
\end{equation}
The parabolic  change of concentration $c(r)$ implies that the stress
$\sigma(r)$ should be  heterogeneously distributed in the contact region.
From a relation of effective stress and concentration [10], we have
\begin{equation}
\sigma^{e}(r) =-\frac{R T}{\nu_{m}} {\rm ln} \frac{c(r)}{c_{0}},
\label{SIGMA-1}
\end{equation}
where $\nu_m$ is the molar volume of the sediment. We
have used here the condition $\sigma^{e}(r)=0$ at $r=L$. Let $\sigma$
be the averaged effective stress, then
\begin{equation}
\pi L^{2} \sigma=\int^{L}_{0} 2 \pi \sigma^{e}(r) r dr. \label{SIGMA-2}
\end{equation}
Combining (\ref{SIGMA-1}) and (\ref{SIGMA-2}), we have
\begin{equation}
\sigma=-\frac{2 R T}{\nu_{m} L^{2}} \int^{L}_{0} r {\rm ln}
[1-\frac{\rho_{s} \dot e \bar d}{4 c_{0} D_{gb} w} (L^{2}-r^{2})] dr.
\end{equation}
Using (\ref{V:DOTE}) and integrating by parts, we have
\begin{equation}
\sigma = -\frac{R T }{\nu_{m}} [ (1-\frac{1}{B L^{2}}) {\rm ln}
(1-B L^{2}) -1],  \label{SIGMA:SOL0}
\end{equation}
where
\begin{equation}
B=\frac{\rho_{s} \dot e \bar d}{4 c_{0} D_{gb} w}.
\end{equation}
By defining a critical effective stress $\sigma_{c}$
(and equivalently a critical creep rate $\dot e_{c}$ ) when $B L^{2}=1$
\begin{equation}
\sigma_{c} \equiv \frac{RT}{\nu_{m}}, \,\,\,\,\,
\dot e_{c} \equiv \frac{4 c_{0} D_{gb}
w}{\rho_{s} L^{2} \bar d},
\label{sigma:typ}
\end{equation}
equation (\ref{SIGMA:SOL0}) can be rewritten as
\begin{equation}
\frac{\sigma}{\sigma_{c}}=[1-(1-\frac{\dot e_{c}}{\dot e}) \ln (1-
\frac{\dot e}{\dot e_{c}})].  \label{SIGMA:SOL}
\end{equation}
From the typical values of $T \sim 300$ K, $R \sim 8.31$ J mol$^{-1}$
K$^{-1}$, and $\nu_{m} \sim 2.6 \times 10^{-5}$ m$^{3}$ mol$^{-1}$ [11],
we can use the definition (\ref{sigma:typ}) to estimate the
typical value of $\sigma_{c}$, which is about $95$ MPa.
Clearly, if $\mid \! \sigma \! \mid \, \ll \sigma_{c}$, we have
\begin{equation}
\dot e=\frac{4 \nu_{m} c_{0} D_{gb} w}{R T \rho_{s} \bar d L^{2}}
\sigma=\frac{16 \nu_{m} c_{0} D_{gb} w}{R T \rho_{s} \bar d^{3}}
\sigma, \label{CREEP:SMALL}
\end{equation}
which is exactly the creep law. Here we have
used $L=\bar d/2$. A different choice of $L=O(\bar d)$ will
only introduce an additional shape factor into the above relation.
Under  upper-crustal stress conditions $\sigma < 100$ MPa, the above
approximation is valid as we expected.
At higher stress states, we can use $\mid \! \sigma \! \mid \, \gg
\sigma_{c}$, then (\ref{SIGMA:SOL}) becomes
\begin{equation}
\dot e=\frac{4 c_{0} D_{gb} w}{\rho_{s} \bar d L^{2}} [1-e^{-\frac{\nu_{m}
\sigma}{R T}}]. \label{CREEP:DER}
\end{equation}
Let $L^{2}=4 \bar d^{2} /\alpha_{s}$, and $\alpha_{s}=O(1)$ is a shape
factor. The above relation (\ref{CREEP:DER}) becomes
\begin{equation}
\dot e=\frac{\alpha_{s} c_{0} D_{gb} w}{\rho_{s} \bar d^{3}} [1-e^{-\frac{\nu_{m}
\sigma}{R T}}], \label{CREEP:GOOD}
\end{equation}
which degenerates into (\ref{CREEP:SMALL}) when
$\nu_{m} \sigma/RT \ll 1$, but (\ref{CREEP:SMALL})
may be inaccurate when $\mid \! \sigma \! \mid \,
\sim \sigma_{c}$. The new compaction relation
(\ref{CREEP:GOOD}) is more accurate and valid
in a more wide range of parameter variations.

Furthermore, the newly derived viscous compaction
law (\ref{CREEP:GOOD}) shows that the strain rate
due to pressure solution is controlled by many
parameters such as grain size ($\bar d$), grain
geometry ($\alpha_{s}$), temperature ($T$),
grain-boundary diffusion coeffient ($D_{gb}$).
This is consistent with Dewers and Hajash's [12]
empirical law derived from a quartz compaction experiment.
However, since the complicated dependence on many
parameters and nonlinear features in (\ref{CREEP:GOOD}),
various simplified version or approximate forms have
been used by many authors in earlier work [13-16].
One common simplification of (\ref{CREEP:GOOD}) is
its linearised form such as (\ref{CREEP-1}) used by
Rutter [7] and Ortoleva [13]. A slightly different
formulation of this compaction law is expressed in terms of
porosity strain versus effective stress (instead of
using the strain rate). Schneider et al. [14] used
a relationship between porosity and effective
stress while Lander and Walderhaug [15]
used an exponential form of intergranular
volume as a function of effective stress. Revil [16]
used a relationship between porosity strain and effective
stress, which includes time $t$ explicitly in his formulation.
However, these different formulations can be transformed
into a relationship similar to (\ref{CREEP:SMALL}) but
such a transformation may depend on the grain packing
structure because of the calculation of porosity strain and porosity.
For simplicity, we will only use the form
(\ref{CREEP:SMALL}) in the rest of the paper.

\subsection{Boundary conditions}

The boundary conditions for the governing equations
are as follows. The bottom boundary at $z=0$ is assumed to be
impermeable
\begin{equation}
u^{s}=u^{l}=0,
\end{equation}
and a top condition at $z=h$ is kinetic
\begin{equation}
\dot{h}=\dot{m}_{s}+u^{s}, \label{eq:ht}
\end{equation}
where $\dot{m}_{s}$ is the sedimentation rate at $z=h$. Also at $z=h$,
\begin{equation}
\phi=\phi_{0}, \s p_e=p_0,
\end{equation}
where $p_0$ is the applied effective pressure at the top of the
porous media, and $\phi_{0}$ is the initial porosity.

\section{Non-dimensionalization}

If a length-scale $d$ is a typical length [9] defined by
\be
d=\{\k{\xi \dot m_s G}{(\rho_s-\rho_l) g}\}^{\k{1}{2}},
\s G=1+\k{4 \eta_0}{3 \xi_0},
\ee
and the effective pressure is
scaled in the following way
\begin{equation}
p=\k{G (p_{e}-p_0)}{(\rho_{s}-\rho_{l}) g d},
\end{equation}
so that $p=O(1)$. Here $G=1+\k{4 \eta_0}{3 \xi_0}$ is the value
at the basin top. Compaction viscosity $\xi$ varies
slowly with temperature as shown below in
equation (\ref{equ:p-us}) where $\bb \ll 1$
and $\kappa \ll 1$, and the medium viscosity $\eta$ also varies
slowly with temperature so that the factor $\k{4 \eta}{3 \xi}$
does not change significantly because the variations of
these two viscosities may cancel in some way as we now mainly
focus on the region where temperature is relative low ($<400$ K).
Therefore, for simplicity, we take $G$ to be constant.
Meanwhile, we  scale $z$ with $d$, $u^{s}$ with
${\dot m}_{s}$, time $t$ with $ d/{\dot m}_{s}$, permeability $k$ with
$k_{0}$, and write
\be
T=T_0+\k{\gamma d}{T_0} \Theta,
\ee
where $\gamma$ is the thermal gradient, and $T_0$ is the
temperature at the basin top. we thus have
\begin{equation}
-\pab{\phi}{t} +\pab{}{z}[(1-\phi) u^{s}]=0,
\label{VC:MASS-1}
\end{equation}
\begin{equation}
\pab{\phi}{t}+ \pab{(\phi u^{l})}{z}=0, \label{PHI:U}
\end{equation}
\begin{equation}
\phi (u^{l}-u^{s})=\v k(\phi) [\pab{p}{z}-(1-\phi) ].
\end{equation}
The viscous relation becomes
\be
p=-(1+\bb \Theta) e^{-\kappa \Theta} \pab{u^s}{z}.
\label{equ:p-us}
\ee
where
\begin{equation}
\lambda=\frac{k_{0} (\rho_{s}-\rho_{l}) g} {\mu {\dot m}_{s}},
\s \bb=\k{\gamma d}{T_0}, \s  \kappa=\frac{E_a \gamma d}{ R T^{2}_{0}}.
\ee
Adding (\ref{VC:MASS-1}) and (\ref{PHI:U}) together and integrating
from the bottom, we have
\be
u^s=-\phi (u^l-u^s)=-u,
\ee
where $u=\phi (u^l-u^s)$ is the Darcy flow velocity. Now we have
\begin{equation}
\pab{\phi}{t} +\pab{}{z}[(1-\phi) u]=0,
\end{equation}
\begin{equation}
u=-\v k(\phi) [\pab{p}{z}-(1-\phi) ].
\end{equation}
The constitutive relation for permeability $k(\phi)$ is nonlinear
and complicated depending on many parameters such as grain geometry,
grain size distribution, materials and even the sedimentary history.
For simplicity without losing the essence of physical mechanism of
pressure solution concerned here, we use a simpler form
\be
k(\phi)=(\k{\phi}{\phi_0})^m,
\ee
where the exponent $m$ is derived from experimental studies. Recently,
Pape et al. [17] suggested that $m=1 \sim 10$ based on fractal modelling
on permeability and extensive experimental studies for 640 core samples.
Considering earlier investigations [2,4,5,17], we here choose a
relative high value, say, $m=8$, which is a typical value for shaly
sediments.

Different relationship of $p$ and $\phi$ or $u$ leads to different compaction
model equations, and thus we have
\begin{equation}
\pab{\phi}{t} = \v \pab{}{z}\{ (1-\phi)
(\k{\phi}{\phi_0})^m [\pab{p}{z}-(1-\phi)] \},
\label{equ-300}
\end{equation}
\begin{equation}
p=\v (1+\bb \Theta) e^{-\kappa \Theta}
\pab{}{z} \{ (\k{\phi}{\phi_0})^m [\pab{p}{z}-(1-\phi) ] \},
\label{equ-400}
\end{equation}
The boundary conditions are
\be
\pab{p}{z}-(1-\phi)=0, \s {\rm at } \s z=0,
\label{bbb-2-0}
\ee
\be
\phi=\phi_{0},  \s
\dot h=\dot m(t) +\v (\k{\phi}{\phi_0})^m [\pab{p}{z}-(1-\phi) ] \,\,
\s {\rm at} \s z=h(t).
\label{bbb-2-1}
\end{equation}
where $\dot m (t)=O(1)$ is a prescribed function of time, which can be taken
to be one for constant sedimentation on top of the porous media.  Obviously,
$\dot m=0$ if there is no further sedimentation and no increasing  loading
on top of the porous media.

For simplicity, we can use  a prescribed linear temperature
profile
\be
\Theta=h(t)-z.
\ee
It is useful for the understanding of the solutions to get an estimate
for $\lambda$ by using values taken from observations [6,7,9].
By using the typical values of $\rho_{l} \sim 10^{3} \, {\rm kg\, m}^{-3},\,
\rho_{s} \sim 2.5 \times 10^{3} \, {\rm kg\, m}^{-3},\,$
$ k_{0} \sim 10^{-15} -\!\!- 10^{-20}\, {\rm m}^{2}, \, \mu \sim 10^{-3}\,
{\rm N\,s\,
m}^{2}, \, \xi  \sim 1 \times 10^{21}$ N s $ {\rm m}^{-2}, $
$\dot m_{s}  \sim 300\, {\rm m\,\, Ma}^{-1}=1 \times 10^{-11}\,
{\rm m\,\, s}^{-1},\, g \=10 {\rm m \,s}^{-2}, \, G \=1, \,
E_a \sim 3 \,{\rm kcal \, mol}^{-1},\,$
and $\gamma=0.03 \,{\rm K\, m}^{-1}$ =(30 K/1000 m);
then $\v \= 0.01 -\!- 1000$, $\bb \=0.1$, $\kappa\=0.2$ and $d \= 1000$ m.

\section{Numerical Simulations and Asymptotic  Analysis}

\subsection{Numerical Results}

The nonlinear diffusion equations have been solved by using
an implicit predictor-corrector method.
A normalized grid parameterized is used to get a rescaled
height variable $Z=z/h(t)$ in a fixed domain, which will
make it easy to compare the results of different times
with different values
of dimensionless parameters in a fixed frame. This transformation
maps the basement of the basin to $Z=0$ and the basin top to $Z=1$.
The calculations were mainly implemented for the time evolutions
in the range of $t=0.5 \sim 10$ since the thickness in the range
of $0.5 {\rm km} \sim 10 {\rm km}$ is the one of interest in the petroleum
industry and in civil engineering.  Numerical results are briefly
presented and explained  below. The comparison with the asymptotic
solutions for equilibrium state will be made in the next section.

The compaction parameter $\v \= 0.01 -\!- 1000$,   which is the ratio
between the permeability and the sedimentation rate,
defines a transition between the slow  compaction ($\lambda << 1$) and fast
compaction ($\lambda >> 1$). As shown in [5], slow compaction is
the compaction in a boundary layer near the basin bottom, and $\phi
\=\phi_0=0.5$, while the more interesting case is the fast compaction
where porosity $\phi$  reduce quickly. However, the effect of temperature
gradient is not included there.   Therefore, we now mainly   investigate
the effect of temperature gradient in the case of fast compaction when
$\v \gg 1$.

Figure 1 provides the viscous compaction profile of porosity
versus the rescaled height $Z=z/h(t)$ at different temperature gradient
$\b=0.05, 0.25, 0.5$ for $\v=100$ and $t=10$.  We can see that
viscous compaction profile  is more or less  parabolic in the top
region. Temperature gradient greatly influence the compaction behavior as
pressure solution proceed, but the thermal effect is only of
secondary importance, which is consistent with previous results [1].
Compared with the case of constant permeability, porosity decreases
much slower in the present case and this in fact implies the increase
of the pore pressure. As the depth increases, the permeability
$k(\phi)=(\phi/\phi_0)^m$ may become very small as $\phi<\phi_0$
for a relative high value of $m$, which will in turn constrain the flow
through the porous media, and consequently the pore fluid in
sediments gets trapped in the lower permeability zone,
resulting the sudden increase of high pore pressure. This can
explain the general occurrance of the high pore pressure in
sedimentary basins.

Figure 2 gives the basin thickness $h(t)$ as a function of time $t$
for different values of $\v=0.1,10,1000$.
It clearly show that the moving boundary $z=h(t)$ increases almost
linearly with time $t$, which implies that $\dot h=const$, but $\dot
h$ is a function of compaction parameter $\v$.

To understand these phenomena and to verify these numerical results,
it would be very helpful if we can find some analytical solutions to
be compared with. However, it is very difficulty to get general
solutions for   equations  (\ref{equ-300})
and (\ref{equ-400})   because these equations are nonlinear with a
moving boundary $h(t)$.  Nevertheless,  it is still possible and
very helpful to find out the equilibrium state and compare
with the full numerical solutions [4,5].

\subsection{Equilibrium State}

To find out the solutions for  the equilibrium state, we must solve
a nonlinear or a pair of nonlinear ordinary differential equations
whose solution can usually implicitly be written in the quadrature  form.
In order to plot out and see the insight of the mechanism, we also
need to solve these ordinary differential equations (ODEs)
numerically although  the solution procedure is
straightforward. However, it is practical  to get the asymptotic
solutions in the explicit form in the   following cases.

For the viscous compaction, the equilibrium state is governed by
\[ \v \pab{}{z}\{ (1-\phi)
(\k{\phi}{\phi_0})^m [\pab{p}{z}-(1-\phi)] \}=0, \]
\begin{equation}
p=\v [1-\b (h-z)] \pab{}{z} \{ (\k{\phi}{\phi_0})^m [\pab{p}{z}-(1-\phi) ] \},
\label{equ-equ}
\ee
where
\be
\b=\kappa-\bb.
\ee
In deriving the equation (\ref{equ-equ}),
we have used the fact that $\bb \ll 1$ and $\kappa \ll 1$ so that
we can linerise the nonlinear factor in equation (\ref{equ:p-us}) by
using  $(1+\bb \Theta) \exp(-\kappa \Theta) \=1-\b (h-z)$.

The integration of the first equation together with the top  boundary
condition leads to
\be
p=[1-\b (h-z)] \pab{}{z}[\k{(\dot m-\dot h)(1-\phi_0)}{1-\phi}].
\ee
Subsituting this expression for $p$ into equation (\ref{equ-equ}) and
integrating once, we obtain
\[
\k{(\dot m-\dot h)(1-\phi_0)}{1-\phi} = 
\v [1-\b (h-z)] (\k{\phi}{\phi_0})^m \] \be \times \{ [1-\b (h-z)] (\dot m-\dot h)(1-\phi_0)
\k{\p^2}{\p z^2} (\k{1}{1-\phi})-(1-\phi) \},
\ee
whose general solution can also be written in a quadrature. However,
two  distinguished limits are more interesting. Clearly, if $\v \ra 0$,
we have
\be
\dot h=\dot m, \s \phi=\phi_0,
\ee
which is the case of no compaction as discussed in the case of poroelastic
compaction. Meanwhile, if  $\v \ra \infty$, we have
\be
[1-\b (h-z)] (\dot m-\dot h)(1-\phi_0)
\k{\p^2}{\p z^2}(\k{1}{1-\phi})-(1-\phi)=0,
\ee
which is non-autonomous and it is difficult to get its general
solution.   However,  we can assume $\b \ll 1$ and  perturb
the above equation in term of  $\b$,
\be
\phi=\phi\0+\b \phi\1+...,
\ee
and the leading order
equation is
\be
(\dot m-\dot h)(1-\phi_0)
\k{\p^2}{\p z^2}(\k{1}{1-\phi\0})-(1-\phi\0)=0,
\ee
which can be rewritten as
\be
(\dot m-\dot h)(1-\phi_0) \psi''-\k{1}{\psi}=0,
\s \psi=\k{1}{1-\phi\0}.
\ee
By using $\psi''=\psi d \psi'/d \psi$ and integrating from $h$ to $z$,
we have
\be
\k{(\dot m-\dot h)(1-\phi_0)}{2} (\psi')^2=\ln \k{\psi}{\psi_0},
\s \psi_0=\k{1}{1-\phi_0}.
\ee
Rearranging the above equation and changing variable
$\psi=\psi_0 \exp(\Psi^2)$, we get
\be
\int_{\psi_0}^{\psi_0 e^{\Psi^2}} \sqrt{\k{2(\dot m-\dot h)}{(1-\phi_0)}}
 e^{\Psi^2} d \Psi =\int_{h}^{z} dz.
\ee
After integration, we have the solution in terms of the original
variables $\phi\0$ and $z$
\be
i [{\rm erf}{\k{i}{1-\phi\0}}-{\rm erf}{\k{i}{1-\phi_0}}]=\sqrt{\k{2(1-\phi_0)}
{\pi (\dot m-\dot h)}} (h-z).
\label{equ-500-500}
\ee
The first order equation is
\be
(\dot m-\dot h)(1-\phi_0)
\k{d^2 \phi\1}{d z^2}+\phi\1=(h-z) \phi\0,
\ee
By using the leading order solution, the solution for $\phi\1$ is simply
\[
\phi\1 \= \phi_0 (h-z)-(1-\phi_0)^2 \sqrt{\k{(1-\phi_0)}{2 (\dot m-\dot h)}}
e^{-\k{1}{(1-\phi_0)^2}} \] \be \times \{ (h-z)^2-2 A^2 [1-\k{\cos\k{z}{A}}{\cos\k{h}{A}}] \}
\label{equ-500-500-1}
\ee
where $A=\sqrt{(\dot m-\dot h)(1-\phi_0)}$.
The comparison of  viscous solutions (\ref{equ-500-500}) and
(\ref{equ-500-500-1}) with the numerical results is shown in Figure 3 in
the top region where the compaction profile is nearly at equilibrium state
for  $\v=1000$ and $t=10$ for two typical thermal gradients $\b=0.1, \, 0.2$.
The agreement verifies the numerical method and the asymptotic
solution procedure.

\subsection{Comparison With Real Data}

The numerical simulations and its comparison with real
data are shown in Figure 4. The solid curve is the
numerical results and real data are depicted by
$\circ$. The real data are the borehole log data
with a total depth of 3700 m in South China Sea. The rescaled height
$Z=z/h(t)$ varies from $0$ to $1$ corresponds to a depth of
3700 m at basement to the ocean floor. In this simulation,
we got best fitted values of $\v=250$, $m=7.3$,
$\bar \beta=0.14 $, and $t=4.3$ (or real time scale
14.2 Ma).

We can see that porosity near the basin top
decrease nearly parabolically with depth, and
the porosity reduction only becomes significant
at the depth  $h-z \sim \Pi$ derived from
solution (\ref{equ-500-500}) when its right
hand $\sqrt{2(1-\phi_0)/\pi (\dot m-\dot h)} (h-z)=O(1)$,
that is
\be
\Pi=d \sqrt{\k{\pi (\dot m-\dot h)}{2 (1-\phi_0)}},
\ee
which is about $980$ m for $\phi_0=0.4$ and $\dot m-\dot h=0.37$.
In other words, pressure solution becomes only significant
at the depths greater than $\Pi$, which is consistent with
the real data.

\section{Discussions}

The present model of pressure solution in sedimentary basins
incorporates the effect of temperature gradient in the frame of
viscous compaction. Based on the  pseudo-steady state
approximations in the grain boundary diffusion process,
we have been able to formulate a new derivation of viscous compaction
relation by using more realistic boundary conditions adjacent
grain contacts. The nondimensional model equations are mainly
controlled by two parameters $\v$, which is the ratio of
hydraulic conductivity to the sedimentation rate, and the thermal
gradient $\b$. Following the similar asymptotic analysis [5],
we have been able to obtain the approximate solutions
for either slow compaction ($\v \ll 1$) or fast compaction
($\v \gg 1$). The more realistic and yet more interesting
case is when small (but realistic) temperature gradient $\b \ll 1$
and $\v \gg 1$, and the equilibrium solution implies a near parabolic
profile of porosity versus depth. Temperature gradient is a very large
factor controlling compaction process, but it is only of second importance
in the sense that it does not influence the parabolic shape of
compaction curves since the shape is mainly characterized by $\v$.
However, for the same value of $\v$ at the same time, the
individual curve of the compaction profile is essentially
described by the thermal gradient.

The numerical simulations and asymptotic analysis have shown
that porosity-depth profile is near parabolic followed by a sudden
switch of nearly uniform porosity because $\lambda (\phi/\phi_0)^m$
may become small (even $\lambda \gg 1$ due to the big exponent $m$)
at sufficiently large depths. In this case, the
porosity profile consists of an upper part
near the surface where the equilibrium is
attained, and a lower part where the
porosity is higher than equilibrium which appears to
correspond accurately to numerical computations. In the near equilibrium
region, the effect of temperature gradient is very distinguished, the
higher the gradient, the quicker the compaction proceeds. On the other
hand, once in the nearly uniform lower region, the porosity is
essentially uniform, the effect of thermal  gradient is not important
and negligible, which is consistent with previous  numerical
simulations [1]. In fact, the permeability becomes so small that
fluid gets trapped below this region, and compaction virtually
stops.\\
\\
{\bf Acknowledgements.}  I would like to thank the referees
for their very helpful comments and very instructive suggestions.
I also would like to than Prof. Andrew C Fowler for his very helpful
direction on viscous compaction.

\section*{References}

\begin{description}
{\small 
\item[1]  C. L. Angevine and D. L. Turcotte,   Porosity reduction
by pressure solution: A theoretical model for quartz arenites, {\em
Geol. Soc. Am. Bull.}, {\bf 94}, 1129-1134(1983).

\item[2]  R. E. Gibson, G. L. England and M. J. L. Hussey, The Theory of
One-dimensional Consolidation of Saturated Clays, I. Finite Non-linear
Consolidation of Thin Homogeneous Layers, {\em Can. Geotech. J.}, {\bf
17}(2)261-273(1967).

\item[3]   D. M. Audet and A. C. Fowler, A Mathematical Model for
Compaction in Sedimentary Basins, {\it Geophys. Jour. Int.},
{\bf 110} (3) 577-590(1992).

\item[4]   A. C. Fowler and X. S. Yang, Fast and Slow Compaction
           in Sedimentary Basins, {\it SIAM Jour. Appl. Math.},
           {\bf 59}(1)365-385(1998).

\item[5]   A. C. Fowler and X. S. Yang, Pressure Solution and
Viscous Compaction in Sedimentary Basins, {\it J. Geophys. Res.},
B {\bf 104}, 12 989-12 997 (1999).

\item[6]   I. Shimuzu,  Kinetics of Pressure Solution Creep in Quartz,
           {\em Tectonophysics}, {\bf 245}(1)121-134(1995).

\item[7]   E. H. Rutter, Pressure Solution in Nature, Theory and
Experiment, {\em J. Geol. Soc. London}, {\bf 140}(4)725-740(1976).

\item[8]   R. Tada, and R. Siever, Pressure solution during diagenesis,
{\em Ann. Rev. Earth Planet. Sci.}, {\bf 17}, 89-118 (1989).

\item[9]   X. S. Yang,  Mathematical Modelling of Compaction and Diagenesis
           in Sedimentary Basins, D.Phil Thesis, Oxford University
           (1997).

\item[10]   A. M. Mullis,  The Role of Silica Precipitation Kinetics in
determining the Rate of Quartz Pressure Solution, {\it J. Geophys.
Res.}, {\bf 96}(7)1007(1991).

\item[11]   M.D. Zoback, R. Apel, J. Baumgartner, M. Brudy,R. Emmermann,
B. Engeser, K. Fuchs,W. Kessels, H. Rischmuller, F. Rummel and
L. Vernik,  Upper-crustal strength inferred from stress
measurements to 6km depth in the KTB borehole, {\it Nature}, {\bf
365}, 633-635 (1993).

\item[12]   T. Dewers and A. Hajash, Rate laws for water-assisted
compaction and stress-induced water-rock interaction in  sandstones,
{\it J. Geophys. Res.},  B{\bf 100}, 13093-112 (1995).

\item[13]  P. Ortoleva,  {\it Geochemical self-organization}, Oxford
University Press, 1994.

\item[14]  F. Schneider, J. L. Potdevin, S. Wolf, \& I. Faille,
Mechanical and chemical compaction model for sedimentary basin
simulators, {\it Tectonophysics},   {\bf 263},  307-317 (1996).

\item[15]  R. H. Lander, O. Walderhaug, Predicting porosity through
simulating sandstone compaction and quartz cementation, {\it
Bull. Amer. Assoc. Petrol. Geol.}, {\bf 83}, 433-449(1999).

\item[16]  A Revil, Pervasive pressure-solution transfer: a poro-visco-plastic
model, {\it Geophys. Res. Lett.}, {\bf 26}, 255-258 (1999).

\item[17]  H. Pape, C. Clauser and J. Iffland, Permeability prediction
based on fractal pore-space geometry, {\it Geophysics},
{\bf 64}, 1447-1460 (1999).

\item[18]   J. E. Smith, The dynamics of shale compaction and evolution in
pore-fluid pressures, {\em Math. Geol.}, {\bf 3}, 239-263(1971).
}
\end{description}

\def\fig#1#2{\begin{figure} \centerline{\includegraphics[width=2.5in,height=2.25in]{#1}}
\caption{\small #2} \end{figure}}

\fig{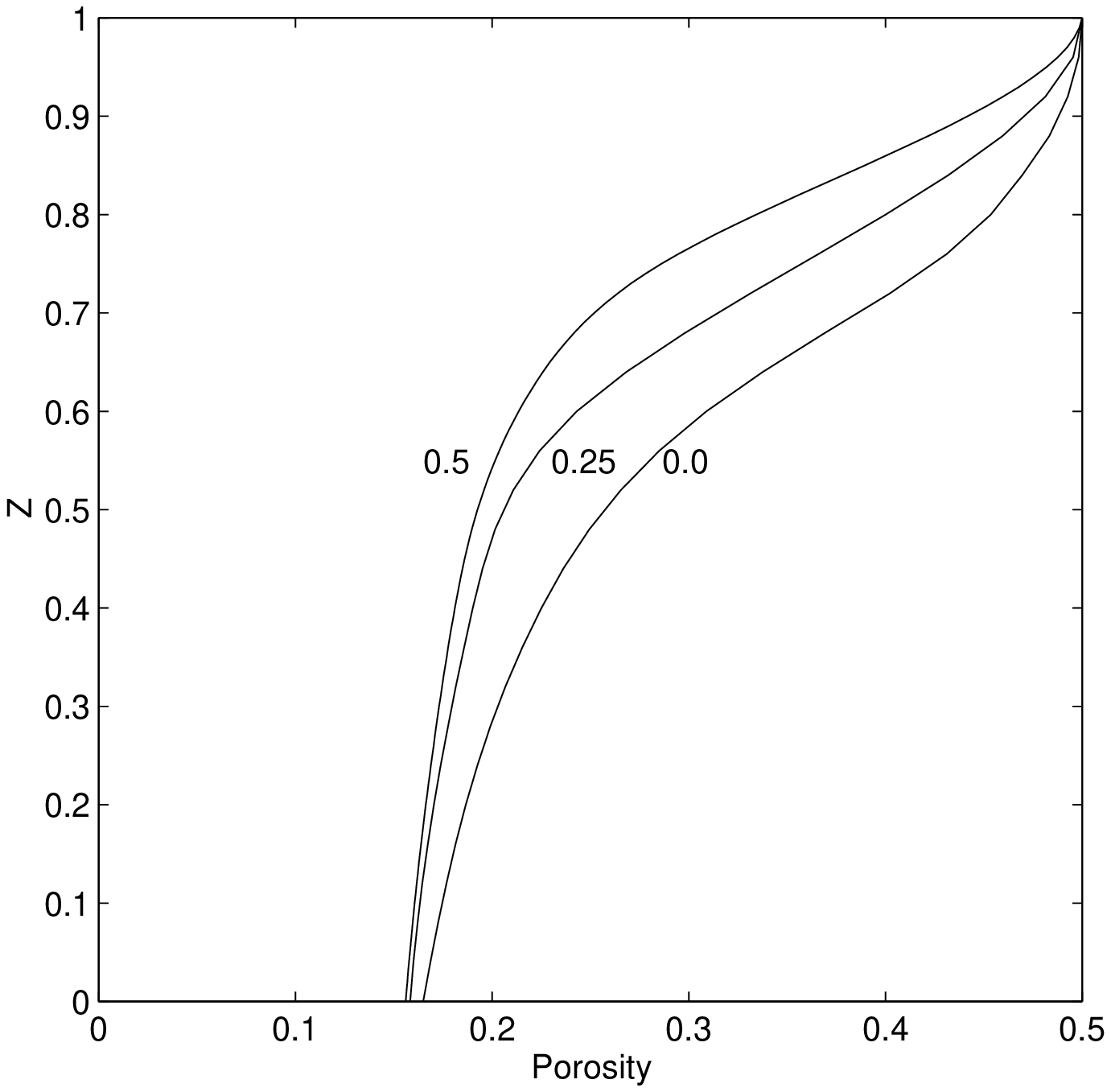}{Viscous compaction profile of porosity versus
          the rescaled height $Z=z/h(t)$ at different temperature gradient
          $\b=0.05, 0.25, 0.5$
          for $\v=100$ and $t=10$. The profile now is nearly parabolic. }
\fig{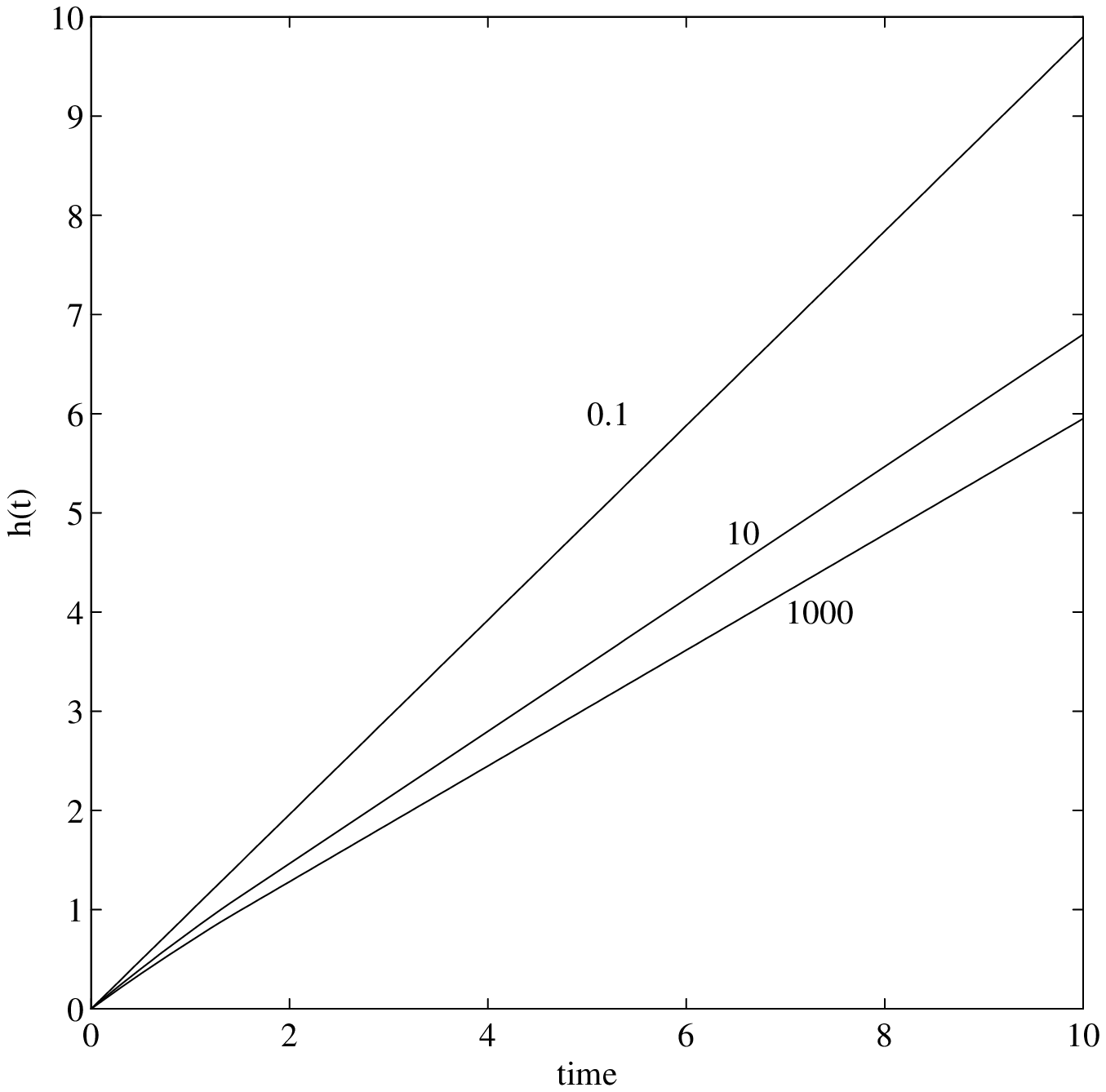}{The basin thickness $h(t)$ as a function of
          time $t$ for different values of $\v=0.1,10,1000$.
          It clearly show that the moving boundary $z=h(t)$ increases almost
          linearly with time $t$, which implies that $\dot h=const$ depending
          only on $\v$. }
\fig{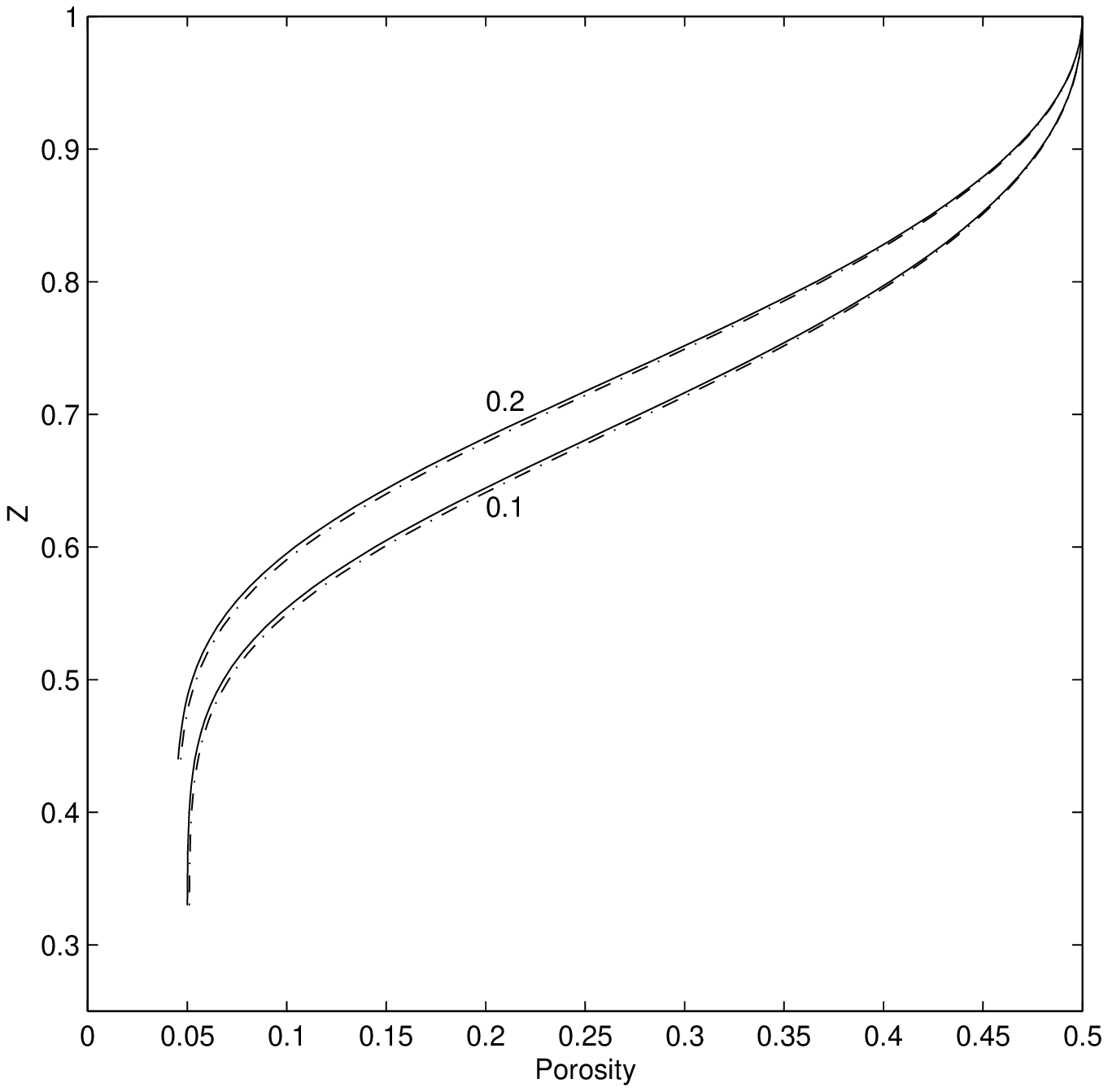}{Comparison of asymptotic solutions  (\ref{equ-500-500})
          and  (\ref{equ-500-500-1})
          (dashed curves) at $t=10$ for $\v=1000$ with  numerical results
          (solid curves) in the
          top region ($Z=z/h(t) \sim 1$ or $z \sim h(t)$)
          where the profile is nearly at equilibrium state.
          The curves are calculated for two typical thermal gradients
          of $\b=0.1, \, 0.2$, and the agreement is    clearly shown.}
\fig{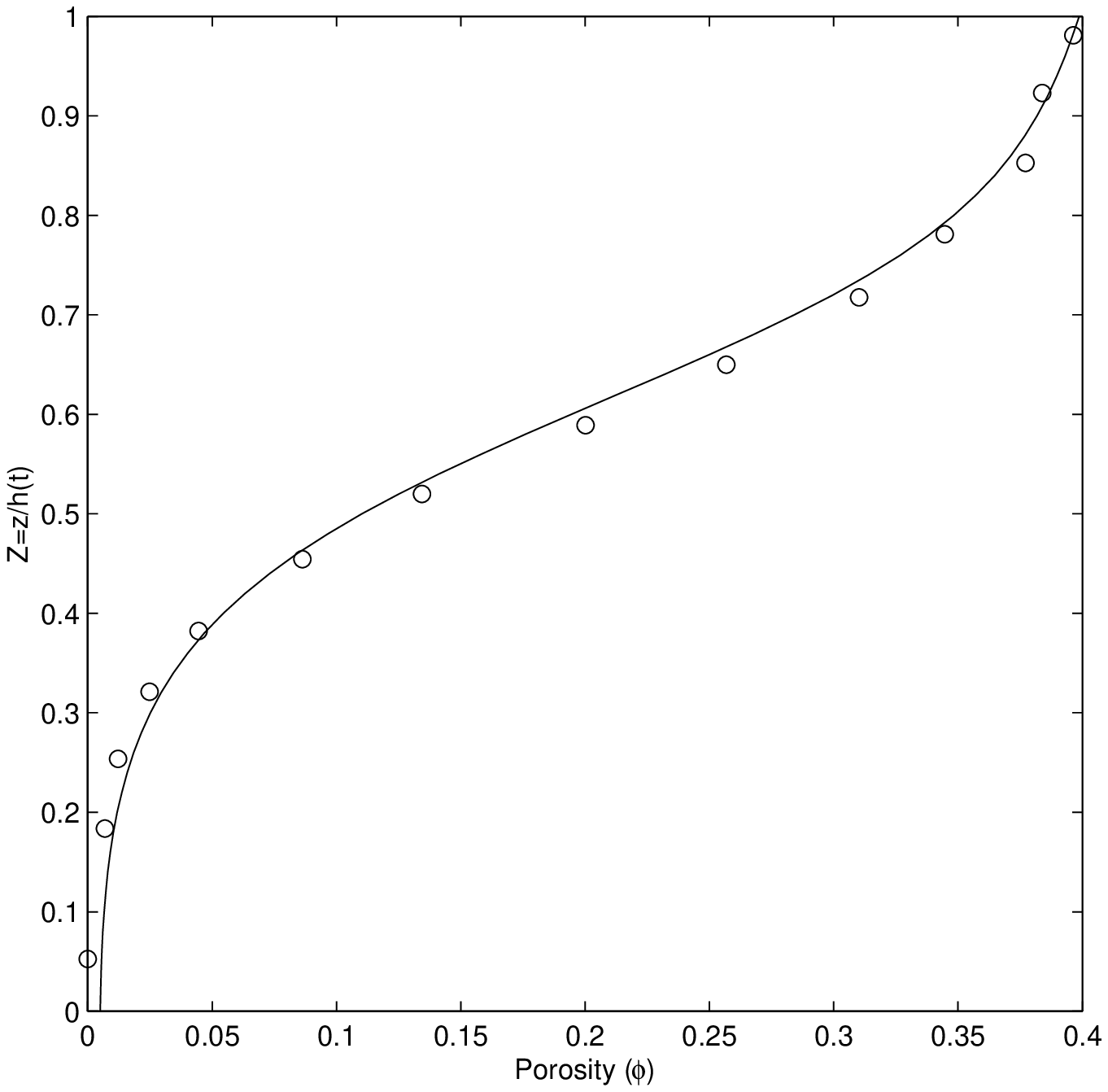}{ Comparison of numerical simulations (solid curves) with
          real borehole log data (with $\circ$).  $Z=z/h(t)$ is
          the scaled height. The best fitted values are $\v=250$,
          $m=7.3$, $\bar \beta=0.14 $, and $t=4.3$.}

\end{document}